\documentclass[fleqn,useAMS,usenatbib]{mnras}
\usepackage{amsmath,subfig}
\usepackage{graphicx,txfonts,fleqn,enumerate,color,hyperref,listings}
\usepackage{ulem} 

\renewcommand*{\v}[1]  {\boldsymbol{#1}}

\newcommand*  {\Exp}[1]  {\mathrm{e}^{#1}}

\newcommand*  {\twovector}[2] {{\begin{pmatrix} $1 \\ $2 \end{pmatrix}}}

\renewcommand {\emph}[1]  {\textit{#1}}

\title[{Reversible integrator switching}]
{\boldmath Switching integrators reversibly in the astrophysical $N$-body problem}

\author[David M. Hernandez and Walter Dehnen]
	{David M.~ Hernandez\href{http://orcid.org/0000-0001-7648-0926}{\includegraphics[width=11pt]{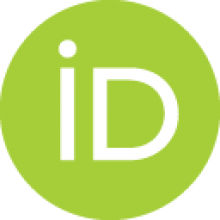}}$^{1}$\thanks{Email: david.m.hernandez@yale.edu} and 
    Walter Dehnen\href{http://orcid.org/0000-0001-8669-2316}{\includegraphics[width=11pt]{orcid-ID}}$^{2,3}$\\ 
	$^1$ Yale University, 52 Hillhouse, New Haven, CT 06511, USA \\
	$^2$ Astronomisches Rechen-Institut, Zentrum f\"ur Astronomie der Universit\"at Heidelberg, M\"onchhofstra\ss{}e 12-14, 69120, Heidelberg, Germany \\
    $^3$ School for Physics and Astronomy, University of Leicester, University Road, LE1 7RH, Leicester, UK
	}
\begin{document}

\maketitle

\label{first page}
\begin{abstract} 
We present a simple algorithm to switch between $N$-body time integrators in a reversible way.  We apply it to planetary systems undergoing arbitrarily close encounters and highly eccentric orbits, but the potential applications are broader. Upgrading an ordinary non-reversible switching integrator to a reversible one is straightforward and introduces no appreciable computational burden in our tests.  Our method checks if the integrator during the time step violates a time-symmetric selection condition and redoes the step if necessary.  In our experiments a few percent of steps would have violated the condition without our corrections.  By eliminating them the algorithm avoids long-term error accumulation, of several orders magnitude in some cases.   
\end{abstract}
\section{Introduction}
\label{sec:intro}
The astrophysical $N$-body problem, with $N$ the number of point masses \citep{sze67,HH03,aar08,DR11},  consists of solving a set of ordinary differential equations that usually describe Newtonian or an approximate relativistic gravity and is required for studying dynamical phenomena ranging from the evolution of planetary systems to formation of dark matter substructure.

The $N$-body problem for $N > 2$ must be solved numerically in many cases.  These calculations run into problems because we do not have access to infinite computing time or infinite computing memory.  The former forces us to search for faster numerical solutions subject to truncation error and the latter results in accumulating error in time due to finite precision.  In this work, we safely neglect finite precision errors which become important when exact trajectories and phases are needed \citep{B37,HMR07,RS15,HH21}.  Eventually the growth of errors in first integrals renders the solution of chaotic problems useless {because the statistics of those solutions are wrong} \citep{Smith77,PB14,BP15,Hernandezetal2020}.  Efforts to control error and improve speed in $N$-body solutions has driven huge progress in astrophysics in the last decades, from simulations of structure formation, to investigations of chaos in the Solar System \citep{S05,LG09,Zeebe2015a,Zeebe2015b}.  

One discovery that has significantly improved the efficiency of $N$-body integrations is symplectic or Hamiltonian integration \citep*{chan90,SC93,leim04,hair06}.  This has become a mainstay in dynamical astronomy despite its limitation of not easily adapting to time and length scales \citep[e.g.,][]{kino91,WH91,sah92,st94,WHT96,LR01,farr07,rein11,Blanesetal2013,Farresetal2013,HB15,Wisdom2018,Petitetal2019,Agoletal2021}, \citep[but note][]{mik99,preto99} .  Time-reversible integration offers a promising alternative in that its error properties are often similar to those of symplectic integrators (\citealt{hair06}; but see \citealt[][]{FHP04,HB18}), but they can more easily adapt to the time and length scales of the problem.  Proposals for exactly reversible N-body integration by \cite{HMM95,Funatoetal96,Makinoetal2006} have remained impractical as they require solving expensive implicit equations at each time step.  As a compromise, researchers have had success with faster, approximately reversible schemes \citep{K00,K98,pelu12,D17}, with adaptive time steps.  A scheme that is as reversible as possible avoids energy drift; the same holds for schemes that are as symplectic as possible.  We will show small irreversibilities lead to large error accumulations over time, just as small breaks in symplecticity can lead to large errors.

Compared to these previous works, in this work we change the strategy to achieve a time-reversible integrator by entirely changing integrators, reversibly, depending on phase space functions, for instance, during close encounters or during pericenter passage.  After the step, we check whether the step met a reversibility condition and redo it if necessary.  In all tests in this paper, $3\%$ or less of steps (often far less) are irreversible and need to be redone.  If not eliminated, this small fraction of irreversible steps would result in significant accumulation of errors.  We present tests that switch between two integrators or steps for simplicity, but a generalization to more switches is straightforward.  

We first present simple tests in which we integrate elliptical orbit solutions from the simple harmonic oscillator and Kepler potentials using switches.  Then we use two realistic, previously studied $N$-body problems to test our algorithm.  First, we study a system that undergoes frequent close encounters, in which we switch integrators during the encounter and can decrease the computational burden of the \texttt{Mercurius}/\texttt{Mercury} integrator \citep{C99,Reinetal2019}.  We also study novel velocity dependent switching functions, which will predict close encounters in \texttt{Mercury/Mercurius}.  Finally, we consider a system in which one of the planets has an eccentric orbit.  We use a Wisdom--Holman method \citep{WH91} with step sizes suggested by \cite{W15,Hetal2022}.  We obtain excellent error behavior even while adapting step sizes.  We also switch integrators during pericenter in this problem and obtain comparable errors to \cite{LD00} at reduced computational complexity.

An outline of the paper follows.  Section \ref{sec:algo} presents the basic reversible algorithm.  Section \ref{sec:first} presents first tests on simple elliptical orbits.  Section \ref{sec:real} studies realistic $N$-body simulations and codes.  We conclude in Section \ref{sec:conc}.
\section{Description of algorithm}
\label{sec:algo}
We aim to solve a conservative mechanical system $\dot{y} = g(y)$, where $y$ is the state vector. Let $h$ be a constant timestep and $M_1^h$, $M_2^h$ two maps for advancing the system from $t_0$ {and $y_0$} to $t_1=t_0+h$ {and $y_1$}. $M_2^h$ is generally more accurate and more expensive than $M_1^h$ (which includes the option of using the same integrator with different step sizes, see Section \ref{sec:adap}). Crucially, $M_2^h$ should be able to resolve time and length scales $M_1^h$ cannot.  For efficient integration, we want to use the cheaper map $M_1^h$ whenever it is deemed sufficiently accurate, as specified by some condition $F(y)>0$ with an appropriate function $F$. A straightforward `naive' method to achieve this is described in Listing \ref{lst:naive}.
\begin{lstlisting}[label={lst:naive},caption={\texttt{Python} code for a `naive' integrator $M_I^h$ switching between the maps $M_1^h$ and $M_2^h$.},
basicstyle=\ttfamily\footnotesize,
breaklines,
language=Python]
01 def mapNaive(y0,M1,M2,F):
02     if F(y0) > 0:
03         return M1(y0)
04     return M2(y0)
\end{lstlisting}
We safely use the terms time-symmetry and time-reversibility interchangeably in this work, but see \cite{hair06}.
This method is not reversible, even if the individual maps $M_1^h$ and $M_2^h$ are, because the condition is checked {(in line \texttt{02}) before} the step, which in reversed time becomes {after the step}. This is a problem, since conservative systems are reversible: switching the sign of velocity changes only the direction but not the path of the solution. Mathematically, if $\hat{\rho}$ {is defined as an operator that} changes the sign of velocities, $\hat \rho g(y)=-g(\hat \rho y)$. As we demonstrate below, irreversible switches from this naive method{, which occur whenever $F(y_0)$ and $F(y_1)$ return different signs,} result in a slow accumulation of errors.

\subsection{A (sufficiently) reversible algorithm}
\label{sec:algo:revers}
To obtain a reversible method and avoid this accumulation of errors, the maps $M_{1,2}^h$ must be reversible and Listing~\ref{lst:naive} must be altered in three ways. First, the condition $F(y_0)>0$ must be replaced by a reversible condition.
{Possible forms are,
\begin{subequations}
\label{eq:conds}
\begin{align}
    \label{eq:CfuncAnd}
    &F(y_0) > 0 \quad\text{and}\quad F(y_1) > 0, \\
    \label{eq:CfuncOr}
    &F(y_0) > 0 \quad\text{or}\quad F(y_1) > 0, \\
    \label{eq:Cfunc}
    &F(y_0) + F(y_1) > 0.
\end{align}
\end{subequations}
Second, $F$ must be constrained to functions}
\begin{align}
   \label{eq:Frev}
{
    F(y) = F(\hat{\rho} y).
}
\end{align}
Not satisfying this relation can lead to poor error performance \citep{HB18}, which we have tested. {Third}, steps which are found to be inconsistent with the conditions \eqref{eq:conds} must be rejected. {For the condition~\eqref{eq:CfuncAnd}}, this leads to {an almost reversible method described} in Listing~\ref{lst:simpleAnd}. {We explain why it is ``almost'' reversible in Section~\ref{sec:algo:exact}.}
\begin{lstlisting}[label={lst:simpleAnd},caption={{
\texttt{Python} code for an almost reversible algorithm that uses condition \eqref{eq:CfuncAnd}}.},
basicstyle=\ttfamily\footnotesize,
breaklines,
language=Python]
01 def mapReversibleAnd(y0,M1,M2,F):
02     if F(y0) > 0:
03         y1 = M1(y0)
04         if F(y1) > 0:
05             return y1
06     return M2(y0)
\end{lstlisting}
{
Listing~\ref{lst:simpleAnd} is identical to Listing~\ref{lst:naive} except for the check of the posterior condition in lines \texttt{04}, \texttt{05}. A slightly more efficient implementation keeping track of $F(y_0)$ (to avoid its repeated computation) is possible. The condition~\eqref{eq:CfuncOr} can be re-phrased as condition $F(y_0) \le 0$ and $F(y_1) \le 0$ for choosing $M_2^h$, which can be implemented equivalently.
}

{
However, in this study we use condition~\eqref{eq:Cfunc}, which is arguably more precise in choosing an algorithm when $F(y)$ changes sign during the step because it interpolates information at the beginning and end of the step. This leads to the algorithm in Listing~\ref{lst:revers:eff}, which keeps track of $F(y_0)$.}

\begin{lstlisting}[label={lst:revers:eff},caption={\texttt{Python} code for an almost reversible algorithm {$M_R^h$ using condition~\eqref{eq:Cfunc}}. This {algorithm (together with} Listing{s}~\ref{lst:simpleAnd} {and~\ref{lst:revers}}{)} is the main result of our study {and is used in all tests in Sections \ref{sec:first} and \ref{sec:merc}.}},
basicstyle=\ttfamily\footnotesize,
breaklines,
language=Python]
01 def mapReversible(y0,F0, M1,M2,F):
02     ytry = M1(y0) if F0 > 0 else M2(y0)
03     Ftry = F(ytry)
04     if (F0 > 0) == (F0 + Ftry > 0):
05         return ytry, Ftry
06     yalt = M2(y0) if F0 > 0 else M1(y0)
07     Falt = F(yalt)
08     if  F0 > 0  or  F0 + Falt > 0:
09         return yalt, Falt
10     return ytry, Ftry
\end{lstlisting}
We denote the algorithm of Listing \ref{lst:revers:eff} as $M_R^h$. {It first attempts in line \texttt{02} a step with the map which is preferred by $F(y_0)$. If the condition~\eqref{eq:Cfunc} agrees with the initial assessment (tested in line \texttt{04}), that step is accepted, which accounts for the vast majority of cases. Otherwise, a step with the other map is attempted (in line \texttt{06}) and accepted if consistent with the condition~\eqref{eq:Cfunc}. Finally, if neither of the attempted steps was consistent, then using $M_2^h$ is chosen (for efficiency and simplicity of code, the last two conditional clauses are combined into one in line \texttt{08}).  In one simple test, the performance difference of Listing \ref{lst:simpleAnd}{, while keeping track of $F(y_0)$,} and Listing \ref{lst:revers:eff} was insignificant, but we have no{t} carried out a comprehensive comparison.}  {It's reasonable to assume that for many problems, the performance difference between the algorithms of Listings \ref{lst:simpleAnd} {(while keeping track of $F(y_0)$)} and \ref{lst:revers:eff} is minor.}  

{In Appendix~\ref{app:algo} we also present an algorithm for a general switch condition combining $y_1$ and $y_0$, which allows, for example, to use the relative change of acceleration over the time step as input for the condition.} Through a series of numerical experiments, we find that $M_R^h$ and $M_I^h$ are equivalent for a majority of time steps, such that converting a naive to a reversible integrator can be done at no appreciable computational penalty.

\subsection{{Why are these methods only almost reversible?}}
\label{sec:algo:exact}
{The algorithms in this section and in Appendix~\ref{app:algo} are,} in fact, not exactly reversible. {The ultimate reason is that the two possible end states $y_1$, $M_1^hy_0$ and $M_2^hy_0$, are not exactly identical and occasionally provide conflicting answers when inserted into the condition~\eqref{eq:Cfunc}. There are to two types of conflicts: inconsistency ($M_1^hy_0$ requires $M_2^h$ but $M_2^hy_0$ requires $M_1^h$, i.e.\ no map is consistent with the condition) or ambiguity (both maps meet the condition). Our algorithm detects inconsistency (and takes $M_2^h$ in this case), but cannot detect ambiguity and uses the map indicated by $F(y_0)$ in this case,} which breaks time symmetry.

{Accounting for ambiguity is possible at significant extra costs, by always attempting $M_1^h$ first and {using  $M_2^h$ only if $M_1^h$} is not consistent with the condition. However, this is still not reversible{, since conflicts may also occur between the forward and backward step}. For example, if $M_1^h$ was not acceptable in the forward direction and hence $M_2^h$ used, a subsequent backward step with $M_1^h$ comes to a state $M_1^{-h}M_2^hy_0$ different from $y_0$, which may be acceptable. Thus, a completely reversible algorithm appears impossible with a condition based on states $y_0$ and $y_1$, unless, perhaps, if one tests the backward step after each forward step.}

{These irreversibilities are caused by differences between the states mapped via $M_1^h$ and $M_2^h$. These differences in turn, are small, proportional to the integration error, such that irreversibilities} are quite rare and do not appreciably affect the long-term error, as we show in Section \ref{sec:longt}. Thus, in this study, we will refer to $M_R^h$ as reversible, even if it is not exactly {so}. 


\section{First tests}
\label{sec:first}
To perform first basic tests, we consider simple elliptical orbit Hamiltonians.  When the origin is at the center, the motion is described by a two-dimensional simple harmonic oscillator Hamiltonian with two equal spring constants in $q_x$ and $q_y$, with two pericenter passages per orbit.  When the origin is at a focus, the Hamiltonian is the Kepler problem, with one pericenter passage per orbit.  We will use $M_2^h$ near pericenters and $M_1^h$ otherwise.

\subsection{Simple Harmonic Oscillator (SHO)}
\label{sec:SHO}
Consider the Hamiltonian for the simple harmonic oscillator in two dimensions with uniform spring constants and period $P = 2 \pi$,
\begin{equation}
H = A + B,
\label{eq:HAB}
\end{equation}
with 
\begin{equation}
A = \tfrac12(p_x^2 + p_y^2),
\label{eq:Asho}
\end{equation}
and
\begin{equation}
B = \tfrac12(q_x^2 + q_y^2).
\label{eq:Bsho}
\end{equation}
A method for solving this is drift-kick symplectic Euler; e.g., $M_{\mathrm{DK}}^h = \mathrm{e}^{h \hat{B}} \mathrm{e}^{h \hat{A}}$.  Here $\hat{B} y = \{y, B\}$, where $y$ is the canonical phase space and $\{\}$ denote Poisson brackets.  A reference for our notation is found in \cite{hair06}.  We derive in Appendix \ref{sec:DKH} the Hamiltonian obeyed by $M_{\mathrm{DK}}^h$.

The popular leapfrog method can be written,
\begin{equation}
M_{\mathrm{DKD}}^h = \mathrm{e}^{\frac{h}{2} \hat{A}} \mathrm{e}^{h \hat{B}} \mathrm{e}^{\frac{h}{2} \hat{A}}.
\label{eq:MDKD}
\end{equation}
In Appendix \ref{sec:DKDH}, we derive the Hamiltonian obeyed by $M_{\mathrm{DKD}}^h$.

\begin{figure}
	\includegraphics[width=90mm]{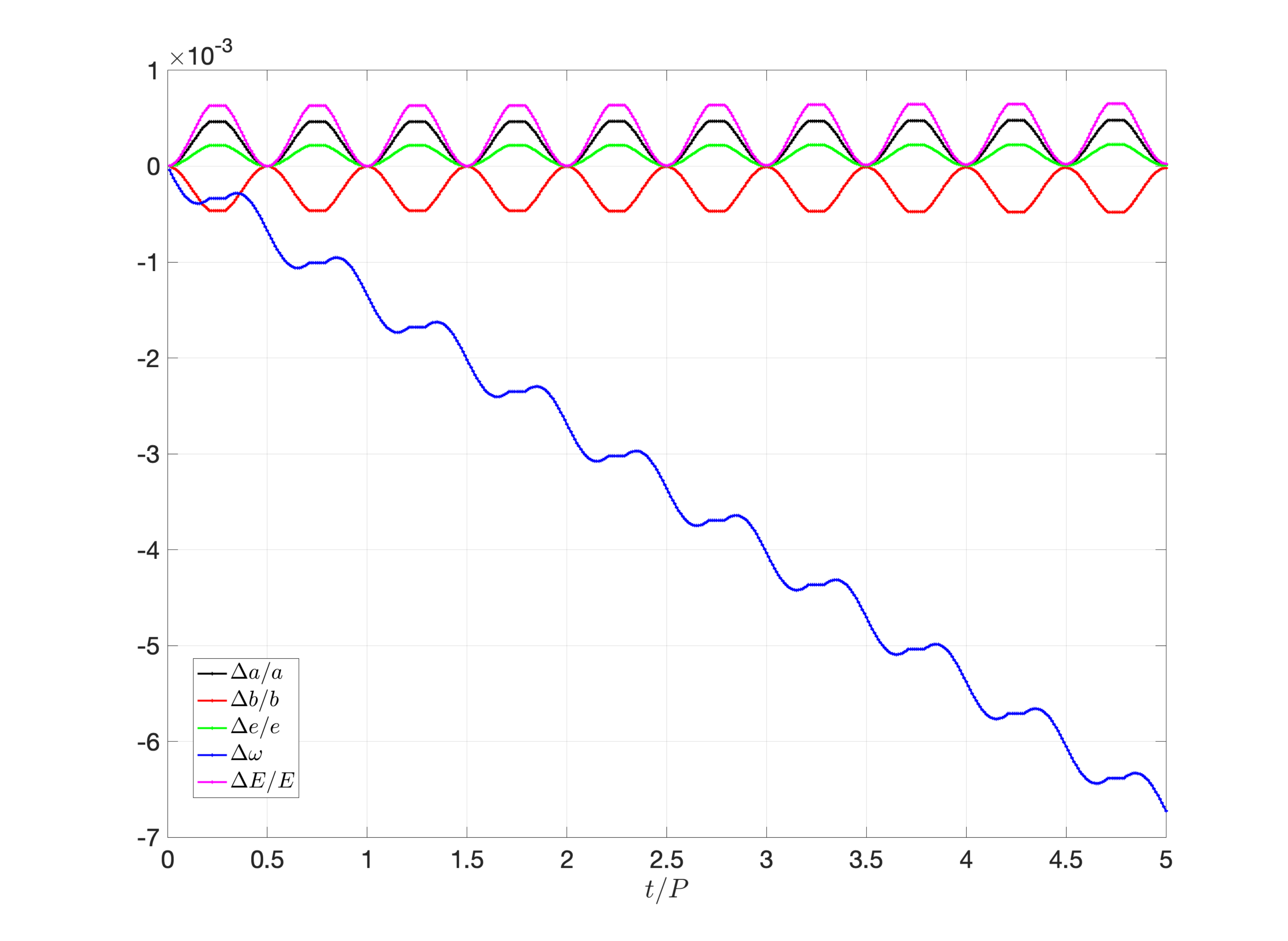}
	
	\caption{Evolution over time of the error in major axis $a$, minor axis $b$, eccentricity $e$, argument of pericenter $\omega$, and energy $E$.  Time is in units of period $P$.  We use $M_R^h$ applied to the 2D simple harmonic oscillator with eccentricity $e = 0.9$.  Orbital parameters remain fixed while $M_2^h$ is used. 
	\label{fig:orbel}
  	}
\end{figure}

Let $M_2^h$ be a map of the exact solution for the two-dimensional SHO and $M_1^h = M_{\mathrm{DKD}}^h$.  Initial conditions are at $(q_x,q_y) = (1,0)$ and $(p_x,p_y) = (0,b)$: rightmost pericenter.  $b$ is the semi-minor axis.  We let the semi-major axis $a = 1$, the eccentricity $e = 0.9$, and $b$ is obtained through, 
\begin{equation}
b = a \sqrt{1 - \mathrm{e}^2}.
\end{equation}
$H = 1-\tfrac12 {e}^2$.  Let $h = P/100$ and $F = \sqrt{\smash[b]{q_x^2 + q_y^2}} - \tfrac12$ in condition~\eqref{eq:Cfunc}.  This criteria forces us to use an expensive exact solution near pericenter.  Our goal for now is not to present optimal switching functions, but to demonstrate the behavior of $M_R^h$.  We compute orbital elements as a function of time.  We assert a solution,
 \begin{equation}
 \begin{aligned}
 \v{q} &= \v{A} \cos (t + \v{\delta}), &
 \v{p} &= -\v{A} \sin(t + \v{\delta}).
 \end{aligned}
 \label{eq:solsho}
 \end{equation}
$\v{A}$ are amplitudes, and $\v{\delta}$ are phases, one of which is $-\omega$, where $\omega$ is the argument of pericenter.  $\omega$ is defined as the angle at which the distance to the center is a minimum, and there is a degeneracy modulo $\pi$ in its definition.  Because the orbits are not closed for $M_R^h$ and $M_I^h$ (Bertrand's theorem does not apply; see \citealt{gol02}), the orbit precesses and we must solve for $\v{\delta}$ as a function of time from $\v{q} \cdot \v{p} = 0$.  Then we obtain semi-major and semi-minor axes, and finally eccentricity.  For $M_R^h$, the evolution over time of the orbital parameters is shown in Fig. \ref{fig:orbel}.
At pericenters, $a$ and $e$ reach local maxima, while $b$ reaches a local minimum.  A maximum precession rate occurs during apocenter.  While $M_2^h$ is used, the orbital parameters are exactly conserved by construction.

\begin{figure}
	\includegraphics[width=90mm]{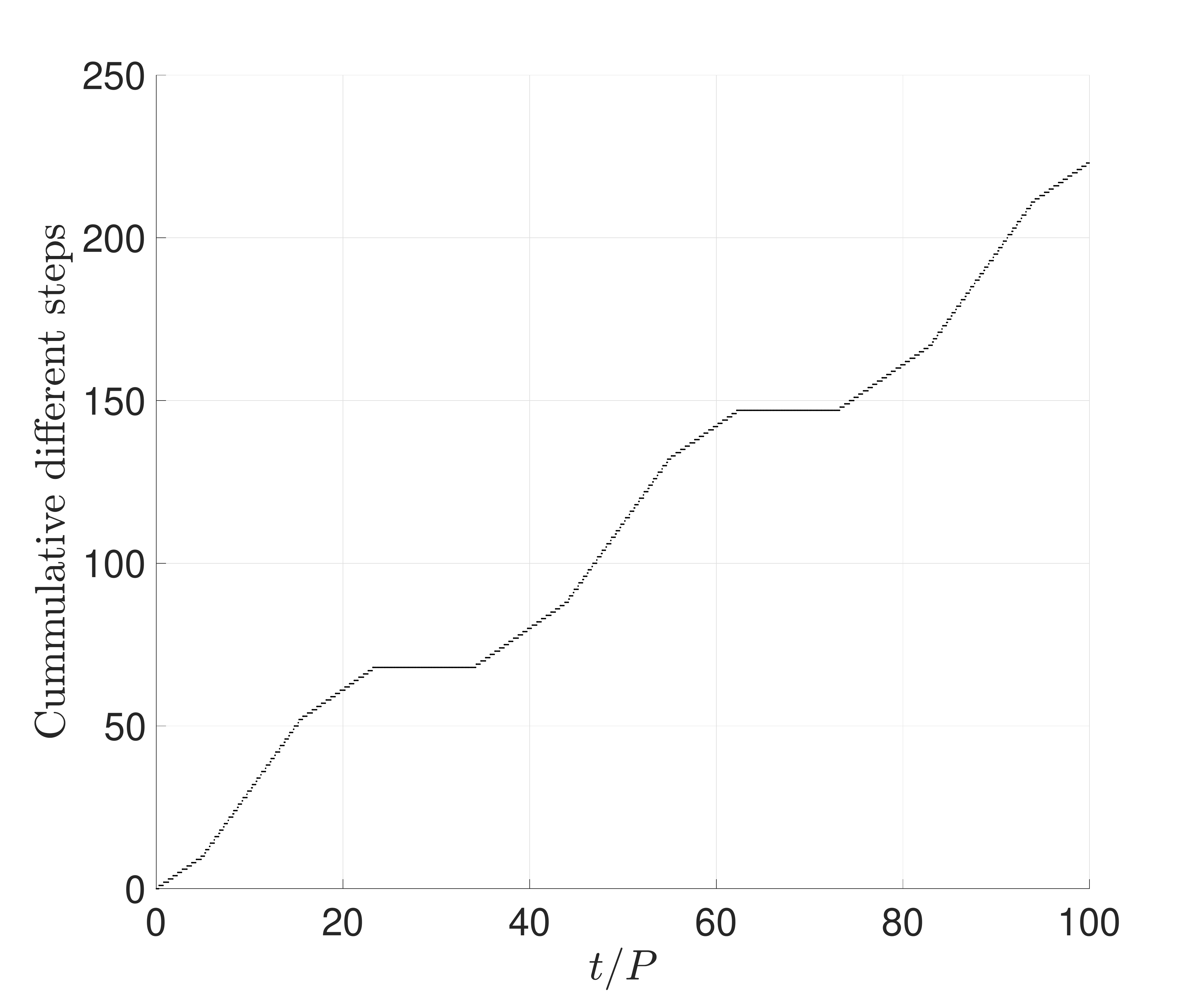}
	
	\caption{ Cumulative number of steps as a function of time for which $M_R^h$ and $M_I^h$ are different for the same problem as Fig. \ref{fig:orbel}.  The slope varies between $0$, $2$, and $4$.
	\label{fig:cumsteps}
  	}
\end{figure}

If $t_i$ are the discrete times at which we apply our maps, we find,
\begin{equation}
M_R^h(t_i) = M_I^h (t_i),
\label{eq:usual}
\end{equation}  
for most $i$.  We illustrate this in Fig. \ref{fig:cumsteps}, which counts the steps for which equation  \eqref{eq:usual} does not hold as a function of $t_i$.  
After $1000P$, this number is $2135$ ($2.135$ different steps per period).  Per period, $M_R^h$ and $M_I^h$ can differ at most four times, the number of switches.  Note the periodic structure of slopes in Fig. \ref{fig:cumsteps}, which persists out to at least $1000P$.  But also note it cannot persist indefinitely as truncation error accumulates, especially for $M_I^h$.  The slopes indicate the number of different steps between $M_R^h$ and $M_I^h$ per period.  During approximately the first five periods, the slope is $2$.  Then it is $4$, and then there is a region of $0$ slope.

\begin{figure}
	\includegraphics[width=90mm]{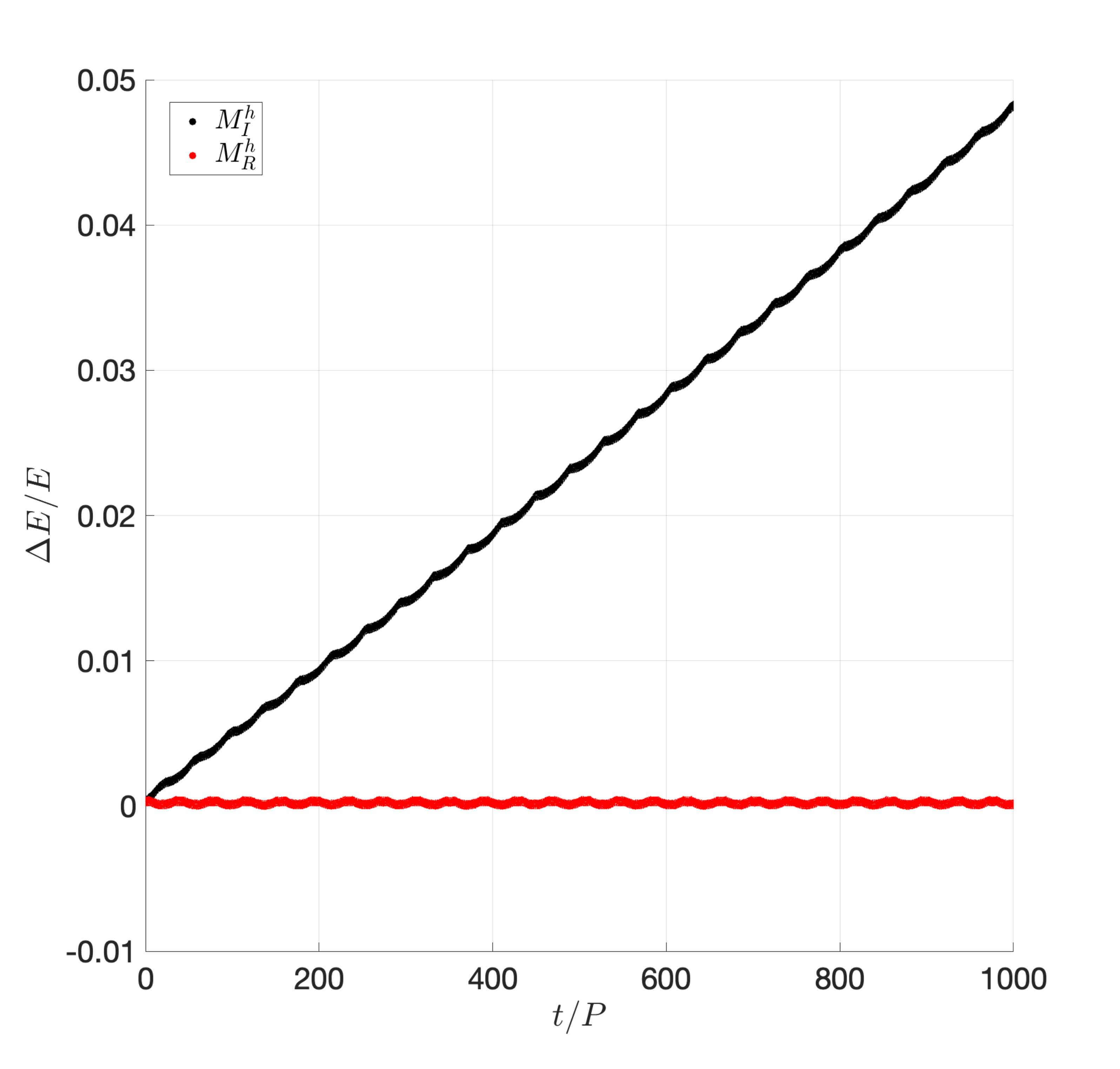}
	
	\caption{Energy error as a function of time for $M_R^h$ and $M_I^h$ applied to the problem of Fig. \ref{fig:orbel}.  The naive integrator yields a linear error drift while the reversible integrator has no clear drift.
	\label{fig:shoecons}
  	}
\end{figure}

We show the energy error for $M_R^h(t_i)$ and $M_I^h(t_i)$ in Fig. \ref{fig:shoecons}.
For $M_I^h$, the error grows linearly in time, while for $M_R^h$ there is no clear error drift.  The naive method introduces an artificial arrow of time, absent in the exact solution, which is responsible for the secular increase in the errors of the integrals of motion.  The final error $\epsilon = \Delta E/E$, with $E = H$, is $\epsilon = 0.049$ for $M_I^h$.  For $M_R^h$, $-2.4 \times10^{-4} < \epsilon < 6.6 \times 10^{-4}$.  Despite the different error behaviors, the cost of $M_R^h$ and $M_I^h$ were essentially the same.  $M_I^h$ had $18011$ calls to $M_2^h$ ($18\%$ of the time) and $81988$ calls to $M_1^h$.  For $M_R^h$, there were $18530$ calls to $M_2^h$ and $83489$ calls to $M_1^h$: these numbers represent an increase by $2.9\%$ and $1.8\%$, respectively.  $2020$ steps were repeated ($2\%$ of total) and no steps were inconsistent.  We checked that no steps were ambiguous over the first $100$ orbits.

\subsubsection{Adapting step size reversibly at no extra cost}
We repeated the experiment of Fig. \ref{fig:shoecons}, replacing $M_2^h$ with $M_2^h$ = $M_{\mathrm{DKD}}^{h/2} M_{\mathrm{DKD}}^{h/2}$: this test will demonstrate $M_R^h$ can be used to adapt a global time step reversibly and discretely at no extra compute cost.  An algorithm with a hierarchy of steps can be designed, as mentioned.  The resulting energy error for $M_R^h$ and $M_I^h$ again shows no secular drift and a linear drift, respectively.  There is no {appreciable} difference in compute cost between $M_R^h$ and $M_I^h$.

\subsection{Kepler Problem}
\label{sec:kepler}
Let the two-dimensional Kepler Hamiltonian have form \eqref{eq:HAB}, with equation \eqref{eq:Asho} and $B = -(q_x^2 + q_y^2)^{-1/2}$.  Let $a = 1$.  Then, $H = -\tfrac12$ and $P = 2 \pi$.  There is no closed form solution in time as in equation \eqref{eq:solsho}.  $M_1^h$ is DKD leapfrog, equation \eqref{eq:MDKD}, and $M_2^h$ is a Kepler advancer.  We use a version adapted from \cite{WH15} to the Matlab programming language. 
{$F = \sqrt{\smash[b]{q_x^2 + q_y^2}} - \tfrac32$ satisfies the reversibility requirement~\eqref{eq:Frev}.}
The computational cost of $M_1^h$ relative to $M_2^h$ depends on implementation, $h$, and initial conditions of the Kepler orbit.  Roughly, we define a cost, 
\begin{equation}
\label{eq:cost}
c = 0.21 N_1 + N_2, 
\end{equation}
where $N_k$ is the number of calls to $M_k^h$ {and the factor} {$0.21$ is determined numerically and empirically.}  We set $e = 0.9$ and $h = P/100$ again.  $(q_x,q_y) = (1+e,0)$, and $(p_x,p_y) = (0, \sqrt{(1-e)/(1+e)})$, at apocenter.  For $M_R^h$, roughly $58\%$ of the steps initially use $M_1^h$. {This fraction is expected to change as truncation error accumulates over a large number of periods.} Comparing $M_R^h$ and $M_I^h$, we get error drifts that look as in Fig. \ref{fig:shoecons}.

\begin{figure*}
	\includegraphics[width=\textwidth]{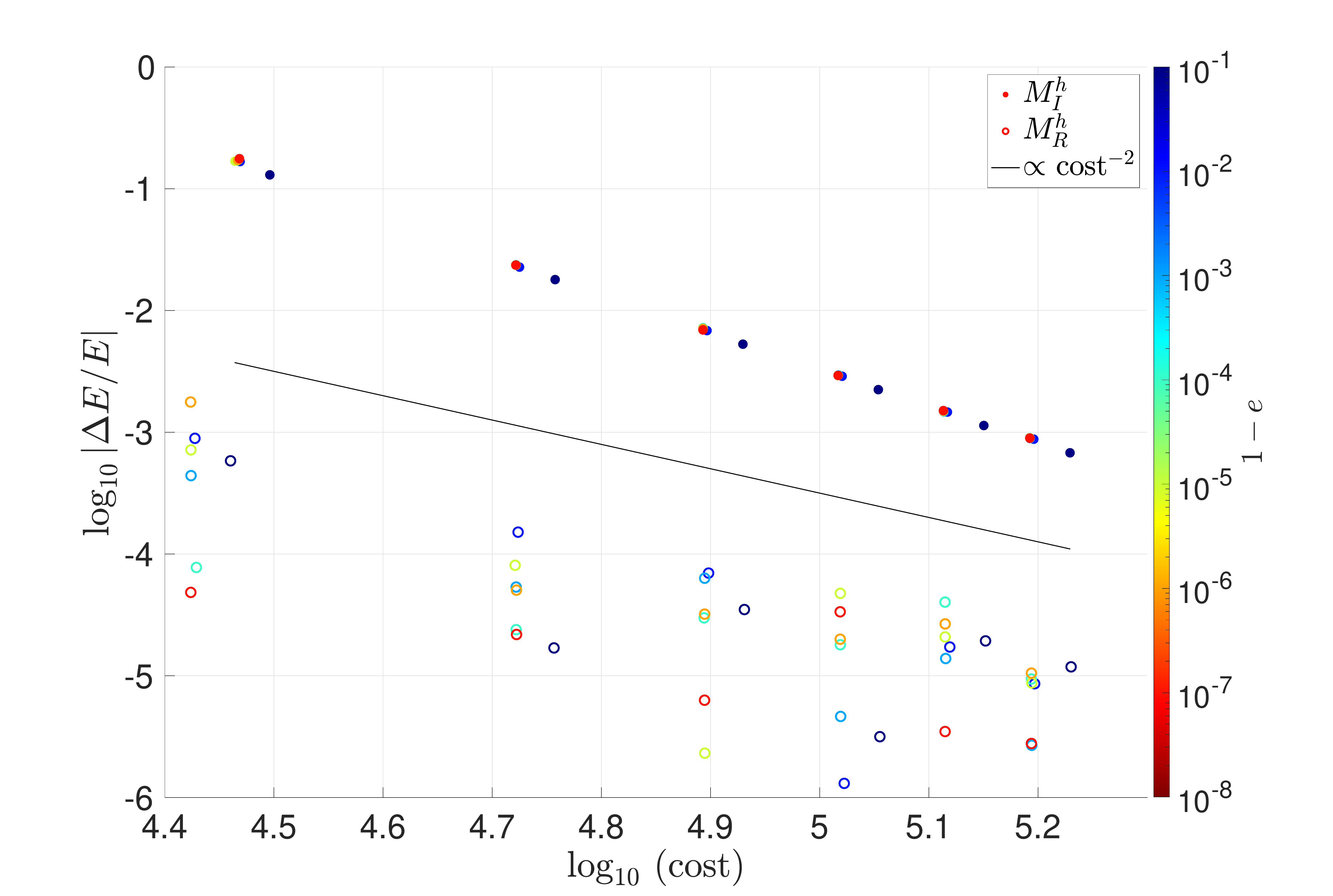}
	\caption{Efficiency plot of the cost versus {absolute value of the cumulative} energy error for integrations of the Kepler problem spanning $1000P$, where $P$ is its period.  The cost measures computational burden and is described {by equation \eqref{eq:cost}}. The step size is varied from $P/50$ to $P/200$ and one minus the eccentricity from $10^{-1}$ to $10^{-7}$. Errors are plotted for the reversible integration $M_R^h$ and naive integration $M_I^h$.  The errors of $M_R^h$ are smaller by roughly two orders of magnitude.
	\label{fig:effic}
  	}
\end{figure*}

Next, we would like to plot an efficiency figure, which compares a wide range of eccentricities and step sizes.  We use six different $h$ ($h/P=50,\,100,\,150,\,200,\,250,\,300$) and seven eccentricities ($1-e = 10^{-1},\,10^{-2},\, 10^{-3},\,10^{-4},\,10^{-5},\, 10^{-6},\, 10^{-7}$), and calculate the absolute value of the cumulative energy error after $1000 P$ as a function of the cost, equation \eqref{eq:cost}, using both $M_R^h$ and $M_I^h$.  For the $42$ integrations with $M_R^h$, between $97\%$ and $99\%$ of steps were not rejected. The fraction of inconsistent steps (see Section \ref{sec:algo:revers}) was between $0$ and $4 \times 10^{-5}$.  The result is shown in Fig. \ref{fig:effic}.
$M_R^h$ yields errors which are consistently about two orders magnitude smaller than $M_I^h$ without computational penalty. The error scales as $c^{-2}$, but for $M_I^h$ the scaling breaks down at high errors where truncation error no longer dominates. Note that using $M_1^h$ alone to solve this problem usually fails.

\subsubsection{Long-term tests}
\label{sec:longt}
\begin{figure}
	\includegraphics[width=90mm]{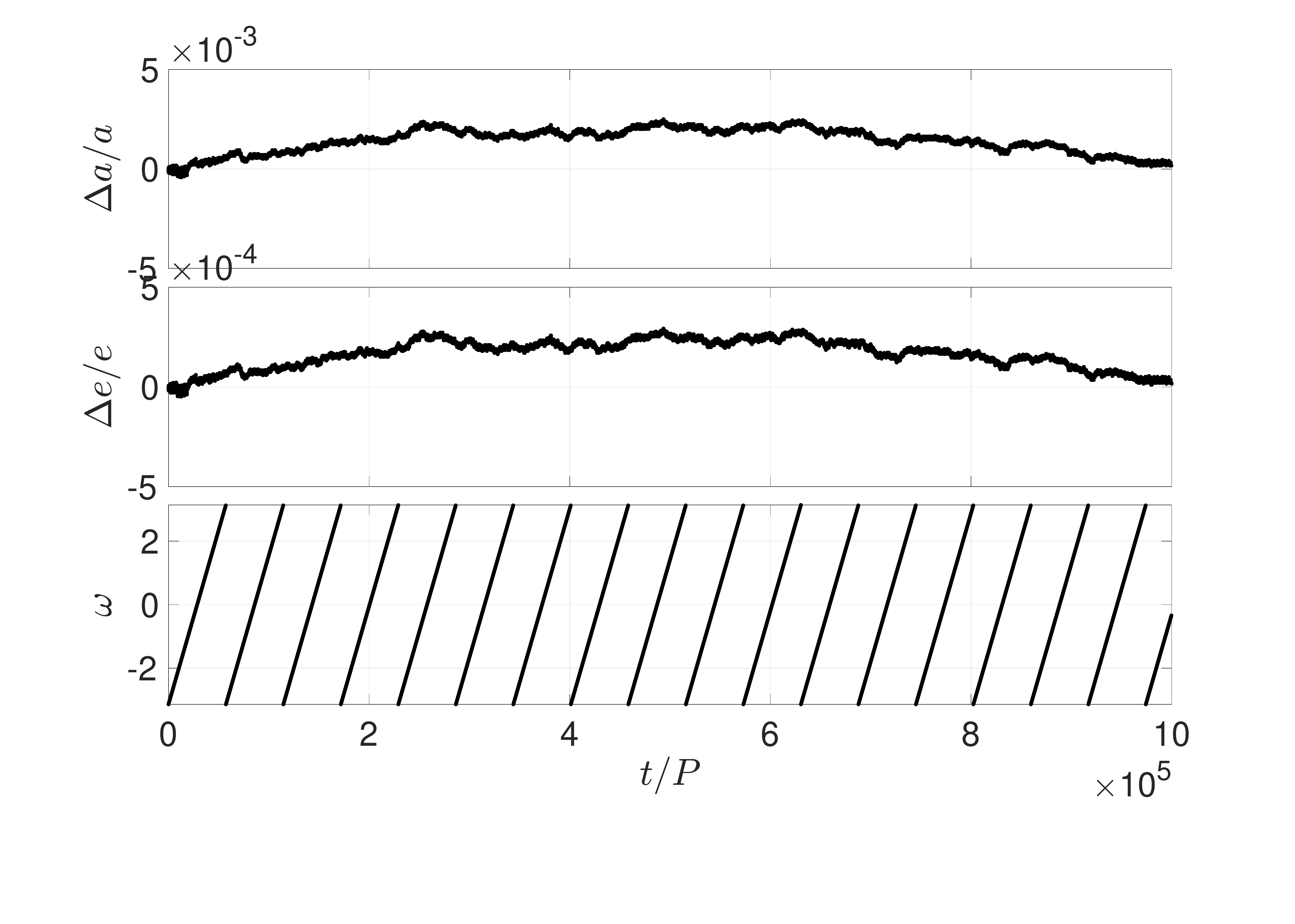}
	\caption{ Long-term error in orbital elements for the Kepler problem with $e = 0.9$ and $h = P/100$.  We have used the reversible $M_R^h$.  Its errors are stable while the errors of $M_I^h$ quickly grow outside the bounds of the figure, except in the case for $\omega$.
	\label{fig:longt}
  	}
\end{figure}

Our tests so far have been limited to $10^3$ orbits, but we would like to explore any long-term error drift in $M_R^h$. To study this, we implement $M_R^h$ for the Kepler problem, as defined in Section \ref{sec:kepler}, in the \texttt{C} programming language.  We reuse initial conditions and parameters from Section \ref{sec:kepler} ($e = 0.9$ and $h = P/100$), and integrate for a million orbits. The error in orbital elements for $M_R^h(t_i)$ is shown in Fig. \ref{fig:longt}.  
For $M_I^h$, the errors in $a$ and $e$ quickly grow outside the bounds of the figure and are, at the final step, $-0.20$ and $-0.03$, respectively.  $\omega$ gets stuck at $-1.11$ radians before $5 \times 10^4 P$.  These errors saturate because after time $t/P = 25027$, $F<0$ for all steps, meaning an exact solution is used for the remaining integration. For $M_R^h$, the error in $a$ and $e$ is far more controlled.  The pattern for the errors in $a$ and $e$ track each other as they did for the SHO in Fig. \ref{fig:orbel}.  $\omega$ circulates about $17$ times over the million orbits.  Of course, one could also solve this problem using $M_2^h$ directly, at higher expense.  

In this test, we found {$122$ ambiguous $(1.2\times10^{-4} \%)$ and $103$ inconsistent $(1.0 \times 10^{-4} \%)$} steps.  If we default to $M_2^h$ during an ambiguous step, there is no qualitative impact on the trend of the error behavior. {We have also checked for irreversible steps. To do this, after each step, we integrate backwards a step and see if the forwards and backwards integrators agree. We can also count the number of ambiguous and inconsistent steps in the backwards direction. We find there are $216$ irreversible steps. There are $213$ and $0$ ambiguous and inconsistent steps, respectively, in the backwards direction.}

The existence of irreversible steps and ambiguous and inconsistent steps{, both in the forwards and backwards directions,} is due to integration error. {The integration error order is given by the error of $M_1^h$, the least accurate method.}  We therefore expect the number of each to scale as $h^2$. {Indeed, as we decrease the time step, we find the number of each kind of step decreases consistent with this scaling.} The fraction of steps in one of these cases scales as $h^3$. {$1011567 (1.0\%)$} steps were redone. The fraction of repeated steps scales as $h$, which we also verified.

\section{\boldmath Realistic astrophysical $N$-body tests}
\label{sec:real}
Having demonstrated the power of the reversibility algorithm on simple problems, we can readily apply these ideas to popular codes solving the $N$-body problem in astrophysics.

\subsection{{Integrating close encounters in planetary simulations}}
\label{sec:merc}
\texttt{Mercury} and \texttt{Mercurius} are a hybrid symplectic integrator for planetary dynamics with a dominant mass that switches to more accurate methods during close encounters of non-Solar bodies.  Works that study this code include \cite{C99,Reinetal2019}.  Briefly, the integrator is,
\begin{equation} 
\label{eq:integ}
M_{\mathrm{merc}}^h = \Exp{\frac{h}{2} \hat{B}} \Exp{h \hat{A}}{}\Exp{\frac{h}{2} \hat{B}}{}.
\end{equation}
We define here,
\begin{equation}
\label{eq:MB}
B =  \frac{1}{2 m_0} \left(\sum_{i \ne 0} \v{P}_i \right)^2 -\sum_{0<i<j} \frac{G m_i m_j}{Q_{i j} } K(Q_{i j}), 
\end{equation}
and
\begin{equation}
\label{eq:MA}
A =  \sum_{i \ne 0} \left(\frac{P_i^2}{2 m_i} - \frac{G m_0 m_i}{Q_i} \right) -\sum_{0<i<j} \frac{G m_i m_j}{Q_{i j} } \left( 1 - K(Q_{i j}) \right).    
\end{equation}
$(\v{Q},\v{P})$ are Democratic Heliocentric coordinates \citep[e.g.,][]{DLL98,HD17}.  $\v{Q}_{i j} = \v{Q}_i - \v{Q}_j$.  $G$ is the gravitational constant, and $m_i$ are masses.  $m_0$ is the dominant mass.  $K$ is called a switching function.

The equations of motion for function $A$ are \citep{Reinetal2019},
\begin{eqnarray}
    \v{  \dot Q}_i &=& \v V_i \label{eq:qdot1}\\
    \v{  \dot V}_i &=& - \frac{Gm_0}{Q_i^3} \v Q_i \nonumber\\
             && - G \sum_{j\neq i, j>0} \frac{ m_j}{Q_{ij}^3} \v Q_{ij} \left(1-K(Q_{ij}) + Q_{ij}  K'(Q_{ij})\right),\label{eq:eomhkh}
\end{eqnarray}
where $\v{V}_i = \v{P}_i/m_i$.
The equations of motion for function $B$ are,
\begin{eqnarray}
\label{eq:test}
    \v{  \dot Q}_i &=& \frac{1}{m_0} \sum_{j \ne 0} \v{P}_j \label{eq:test1} \\
    \v{\dot V}_i &=&  - G \sum_{j\neq i, j>0} \frac{m_j}{Q_{ij}^3} \v Q_{ij} \left( K(Q_{ij}) 
             -  Q_{ij} K'(Q_{ij}) \right). \label{eq:eomhik}
\end{eqnarray}
\cite{Reinetal2019,H19b} proposed various functions for $K$ and the force switching function $L(x) = K(x) - x K'(x)$.  \cite{H19b} studied their accuracy over long time scales in a planetary system, while \cite{Reinetal2019} studied them over one close encounter.  If $K$ is non-constant, solving equations \eqref{eq:qdot1}, \eqref{eq:eomhkh}, \eqref{eq:test}, and \eqref{eq:eomhik} requires evaluating $K$ and $K'$ which might be expensive.  \cite{Reinetal2019} proposed a modified Heaviside function defined as a function whose value remains unchanged during the step as $0$ or $1$; we apply this idea to $K$ so that it is chosen at the start of the time step and $K' = 0$.  Thus, equations \eqref{eq:eomhkh} and \eqref{eq:eomhik} become far simpler.  We note that with this modified Heaviside function, symplecticity is lost.  But this is a strategy which can be made time-reversible through application of $M_R^h$ from Section \ref{sec:algo}.  

\begin{figure*}
	\includegraphics[width=\textwidth]{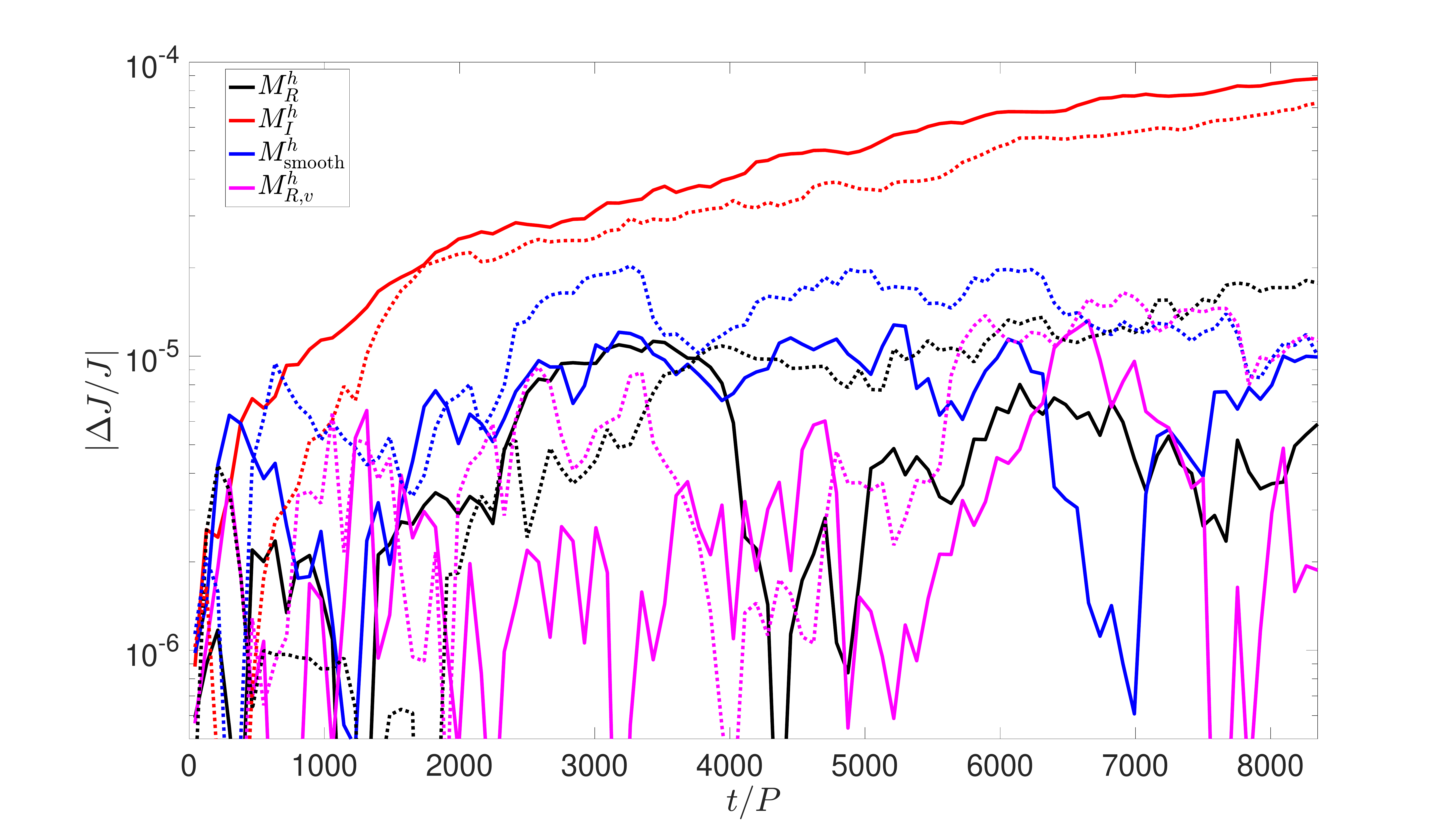}
	
	\caption{Error over time for integrations of a chaotic exchange orbit of the restricted three-body problem.  The error for a second pair of initial conditions, obtained by displacing the $x$-coordinate of the test particle, is indicated with a dotted line.  The error using $M_R^h$, $M_I^h$, and a symplectic $M_{\mathrm{smooth}}^h$ are plotted.  The three maps are obtained by adapting the \texttt{Mercury}/\texttt{Mercurius} integrator.  $M_R^h$ is computationally far simpler than $M_{\mathrm{smooth}}^h$, but their errors are comparable. {We also {used} 
    {$M_{R}^h$} with a velocity dependent switching function {(magenta, labelled $M_{R,v}^h$)}. 
    {For} the same initial conditions and parameters {we find it comparable to the position-dependent switch}.
    }
	\label{fig:mercurius}
  	}
\end{figure*}

To test this idea, we consider a chaotic exchange orbit in the planar restricted three-body problem. This problem has been studied thoroughly in previous works \citep{Wisdom2017,DH17,H19b}: the test particle executes many close encounters with the secondary, which we call ``Jupiter'', a situation for which \texttt{Mercury} and \texttt{Mercurius} were designed.  The test particle does not get arbitrary close to the primary, even after $100,000$ years when solving the exact equations of motion, presumably due to the existence of a conserved quantity (see trajectory plots in \cite{Wisdom2017}.  We use the implementation of Mercurius as described in \cite{H19b}.  We define $F = Q_{23} - R_{\mathrm{H}} a$, where $R_{\mathrm{H}}$ is the Hill radius of Jupiter and $a = 4$.  If $F < 0$, $K = 0$ and if $F > 0$, $K = 1$. 
$h = 8$ days, and the error in the Jacobi constant of the test-particle over time is shown in Fig. \ref{fig:mercurius}.
$P$ is Jupiter's period ($= 11.86$ yrs).  A second pair of initial conditions is also plotted by displacing the test particle $x$-coordinate by $+0.01$ au.  Output was produced every year and a median error every $1000$ years is plotted in Fig. \ref{fig:mercurius}.

$M_R^h$ improves the absolute error over $M_I^h$ by a factor of $11$ to $4$ at the final time.  Over the first $1000$ years, $M_R^h$ and $M_I^h$ differed only in $0.1\%$ of steps.  We also plot $M_ {\mathrm{smooth}}$, which solves equations \eqref{eq:eomhkh} and \eqref{eq:eomhik} with $K(x) = 6 x^5 - 15 x^4 + 10x^3$ for $0 \le x \le 1$.  For $x < 0$, $K(x) = 0$ and for $x > 0$, $K(x) = 1$.  $x$ is defined in \cite{H19b} and all parameters are reused from that work.  $K$ is a $C^2$ function, so $M_{\mathrm{smooth}}^h$ has the same smoothness as the integrator by \cite{Wisdom2017}.  $M_{\mathrm{smooth}}^h$ is symplectic according to \cite{H19}.  It is more computationally complex to solve than $M_R^h$ and $M_I^h$, although we refrain from reporting timing costs as we are using an interpretive programming language to carry out these tests.  We see in Fig. \ref{fig:mercurius} that the far simpler, non-symplectic $M_R^h$ matches the accuracy of $M_{\mathrm{smooth}}^h$.  Thus, for this problem, we have accomplished a stated goal of \cite{Reinetal2019} to find a computationally simple algorithm (and likely faster, depending on implementation) which matches the accuracy of symplectic schemes. \cite{Reinetal2019} used a modified Heaviside function in their algorithm which was irreversible, which used infrequently, as we show here, can yield adequate results.

\subsubsection{\texttt{Mercury/Mercurius} with velocity dependent switching functions}
We can write a version of \texttt{Mercury/Mercurius} that accounts for particles with large relative velocities such that they might have a close encounter on a timescale short compared to the time step.  To our knowledge, velocity dependent switching functions for these codes have not been considered previously.  A velocity dependent switching function causes \eqref{eq:MB} to be non-integrable, but our discrete switching, in which the $K$ can only take two values, bypasses this problem.  Write the Hamiltonians as, 
\begin{equation}
\label{eq:MBv}
B =  \frac{1}{2 m_0} \left(\sum_{i \ne 0} \v{P}_i \right)^2 -\sum_{0<i<j} \frac{G m_i m_j}{Q_{i j} } K(Q_{i j},V_{ij}), 
\end{equation}
and
\begin{equation}
\label{eq:MAv}
A =  \sum_{i \ne 0} \left(\frac{P_i^2}{2 m_i} - \frac{G m_0 m_i}{Q_i} \right) -\sum_{0<i<j} \frac{G m_i m_j}{Q_{i j} } \left( 1 - K(Q_{i j},V_{ij}) \right).
\end{equation}
Inspired by \cite{Antonanaetal2021} (but see also \citealt{Handsetal2019,Boekholtetal2022} ), let 
\begin{equation}
F_{i j} =  Q_{ij}\left(\sqrt{3 V_{ij}^2+ G (m_i + m_j)/Q_{ij}}\right)^{-1} - a h,
\end{equation}
where we have set $a = 30$ somewhat arbitrarily.  $F_{ij}$ takes into account the close encounter and free-fall times.  We construct $M_R^h$ and apply this new integrator to the problem of Section \ref{sec:merc} with the same $h$.  The result is plotted as $M_{R,v}^h$ in Fig. \ref{fig:mercurius}.  $M_{R,v}^h$ achieves a similar accuracy to $M_{\mathrm{smooth}}^h$.  While a velocity dependent switching function did not prove necessary for this problem, there may be other problems for which they would.  The main novelty of this subsection is to show the ease with which we can implement velocity-dependent switch criteria into \texttt{Mercury/Mercurius}.

\subsection{Adaptive stepping for highly eccentric orbits in planetary systems}
\label{sec:adap}
\begin{figure}
	\includegraphics[width=90mm]{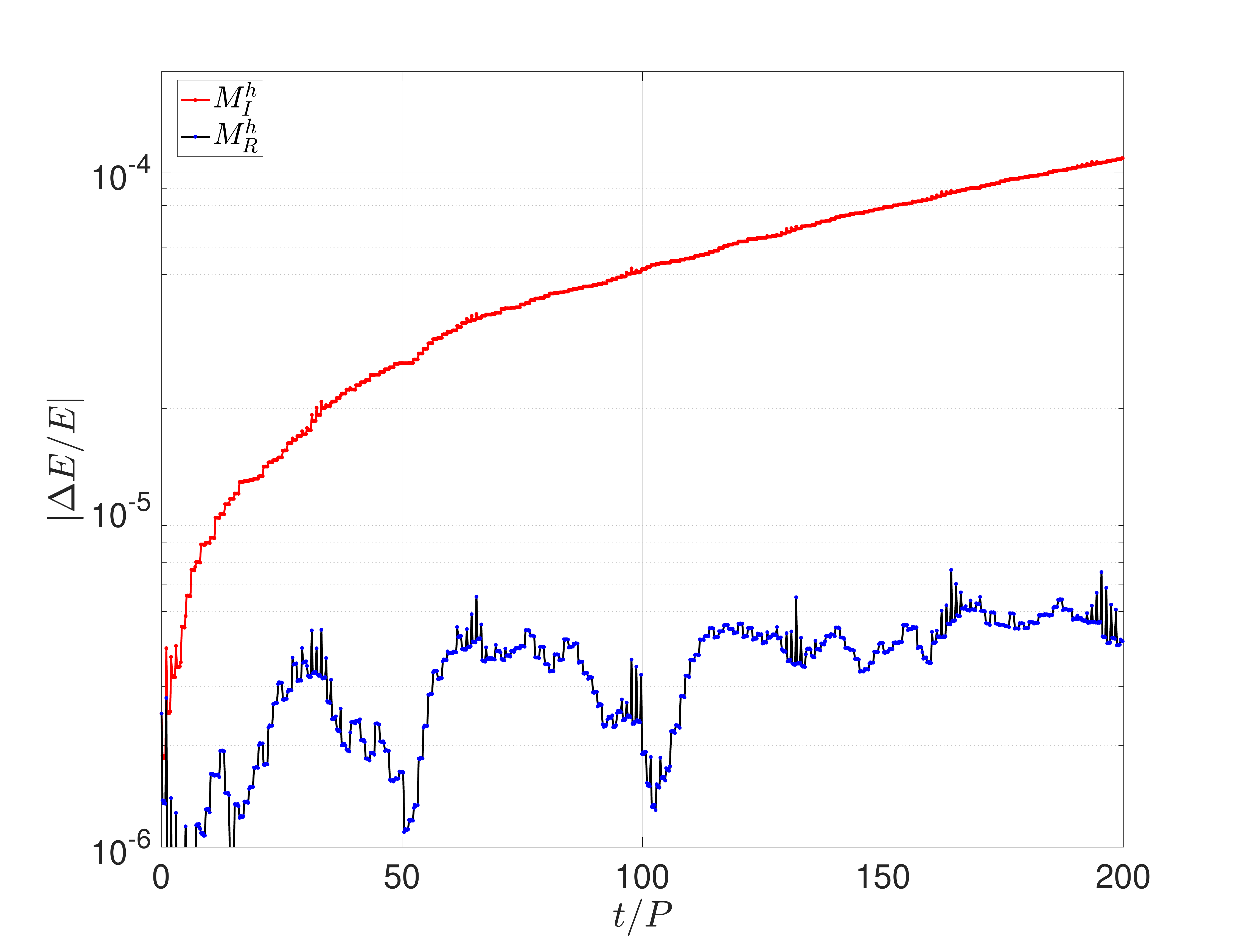}
	
	\caption{The evolution over time, in units of Saturn's period $P$, of the energy error in an $N$-body system consisting of Sun, Jupiter, and Saturn.  Saturn's eccentricity is set to $e = 0.95$ and its inclination to $\pi/2$.  The inclination of Jupiter is $0$.  $M_R^h$ and the naive $M_I^h$ are used to produce these curves.  $M_R^h$ only requires an extra $0.2\%$ of steps compared to $M_I^h$.
	\label{fig:twoplanet}
  	}
\end{figure}

We apply our reversibility scheme to study planetary systems using adaptive steps.  Because it is a well-studied problem, consider the two-planet problem discussed in \cite{LD00,Wisdom2017}.  This planetary system consists of Sun, Jupiter, and Saturn, but Saturn's eccentricity is set to $0.95$ while its inclination is set to $\pi/2$.  Its major axis is left unchanged.  At this eccentricity, Saturn's orbital distance ranges from $0.48$ to $18.9$ au.  We use a mixed-variable symplectic, or Wisdom--Holman \citep{WH91} method in Democratic Heliocentric coordinates to integrate this system for $200$ Saturn periods $P$.  We note that ``mixed-variable symplectic'' is the usual name given, but once we adapt time steps, as in this section, symplecticity is lost.  Saturn reaches a maximum eccentricity of $0.96$ during this time.  The eccentricity increases while inclination decreases.  With these parameters, the results of \cite{Hetal2022} predict that at step sizes greater than $\tau = 0.0015$ yrs, numerical instability occurs (see also \citealt{WH92,RH99}).  Using this small $\tau$ for the entire integration, including near apocenter, is unnnecessarily expensive, but changing $\tau$ would break the symplectic nature of the integrator.  However, we can use our reversibility scheme to ensure reversibility is preserved even if symplecticity is sacrificed, which may work just as well \citep{hair06}.  Thus, we use $h = 6 \tau$ as $M_1^h$ and $6$ steps with step size $h' = h/6$ for $M_2^h$.  More specifically, $M_2^h = (M_1^{h/6})^6 $.  $F = r - 2$ au, where $r$ is Saturn's heliocentric distance.  Fig. \ref{fig:twoplanet} compares the energy error of $M_R^h$ and $M_I^h$ as a function of time.  Output is generated at $1000$ times.
 $2.1\%$ of steps used $M_2^h$ for the $M_I^h$ method.  The $M_R^h$ integration only required redoing $0.2\%$ of steps.  The error of $M_R^h$ is $27$ times smaller than that of $M_I^h$.  This again shows the significant error accumulation caused by a few irreversible steps.  We verified the accumulation of error for $M_I^h$ occurs at pericenter passages.  For $M_R^h$ there is an increase of error at pericenter, but this error decreases after pericenter passage.  This is normal behavior if the time step were to always be kept small at all phases of the orbit, but this is inefficient.
\subsection{Switching integrators near pericenter}
\label{sec:ld}

\texttt{Mercury}, with our modifications in Section \ref{sec:merc} will satisfactorily integrate close encounters between non-Solar bodies, but not close encounters with the Sun. This is because equations \eqref{eq:qdot1} and \eqref{eq:eomhkh}, with $K = K' = 0$ have the wrong gravitating mass {for} non-test particle Hamiltonians.  A solution has been proposed by \cite{LD00}.  We can significantly simplify the approach of \cite{LD00} by moving the terms in \eqref{eq:test1} discretely and reversibly to \eqref{eq:qdot1} for particles undergoing close encounters with the Sun.  The Hamiltonian described by equations \eqref{eq:qdot1} and \eqref{eq:eomhkh} is no longer integrable after such a shift, but the equations for the particle undergoing the close encounter with the Sun can be solved using a high-order method like Bulirsch--Stoer.

In equations, \cite{LD00} propose the mapping,
\begin{equation}
\label{eq:LD}
M_{\mathrm{LD}}^h = \Exp{\frac{h}{2} \hat{B}} \Exp{h \hat{A}}{}\Exp{\frac{h}{2} \hat{B}}{}.
\end{equation}
Now, we have,
\begin{equation}
\label{eq:LDB}
B =  \frac{1}{2 m_0} \left(\sum_{i \ne 0} \v{P}_i \right)^2\left(1 - K(\v{Q})  \right) -\sum_{0<i<j} \frac{G m_i m_j}{Q_{i j} }, 
\end{equation}
and
\begin{equation}
\label{eq:LDA}
A =  \frac{1}{2 m_0} \left(\sum_{i \ne 0} \v{P}_i \right)^2 K(\v{Q} )  + \sum_{i \ne 0} \frac{P_i^2}{2 m_i} - \frac{G m_0 m_i}{Q_i}.    
\end{equation}
$K$ is a smooth switching function as in Section \ref{sec:merc}, but it now depends on heliocentric distances, rather than distances between non-Solar bodes.  If we define, $H_J = 1/(2 m_0) \left(\sum_{i \ne 0} \v{P}_i \right)^2 $, the \cite{LD00} approach consists of smoothly transferring $H_J$ from $B$ to $A$ during a Solar close encounter.  When $K = 0$, both $B$ and $A$ are integrable.  When $K = 1$, during a close encounter, only $A$ is non-integrable and more difficult to solve.  When $0 < K < 1$, both $B$ and $A$ are non-integrable and difficult to solve; this will be a regime the reversible algorithm in our paper completely avoids.  \cite{LD00} show in their Fig. 1 that their proposal, equations \eqref{eq:LD}, \eqref{eq:LDB}, \eqref{eq:LDA}, significantly improves the accuracy of systems with highly eccentric orbits.  We now attempt to simplify and speed up their method while even improving their reported accuracies.  We repeated a more difficult version of the experiment in their Fig. 1.  These initial conditions are a variation of the test of Section \ref{sec:adap}; we now integrate for $3000$ yrs with $h = 0.15$ yrs.  The difference with the \cite{LD00} test is as follows.  In their setup, $K = 0$ for $r < R_1=3$ au, and $r$ is Saturn's heliocentric distance.  $K = 1$ for $r > R_2=4$ au.  For $R_1 < r < R_2$, $0 < K < 1$.  Instead, we use our discrete switching, as opposed to their smooth switching, with $K = 0$ or $K = 1$, determined before the step, according to switching criteria $F = r - 2$ au.  For the reversible algorithm, the step is repeated as needed.  Our algorithm switches at smaller radii than those of \cite{LD00}, making our test more difficult and error prone than theirs.

\begin{figure}
	\includegraphics[width=90mm]{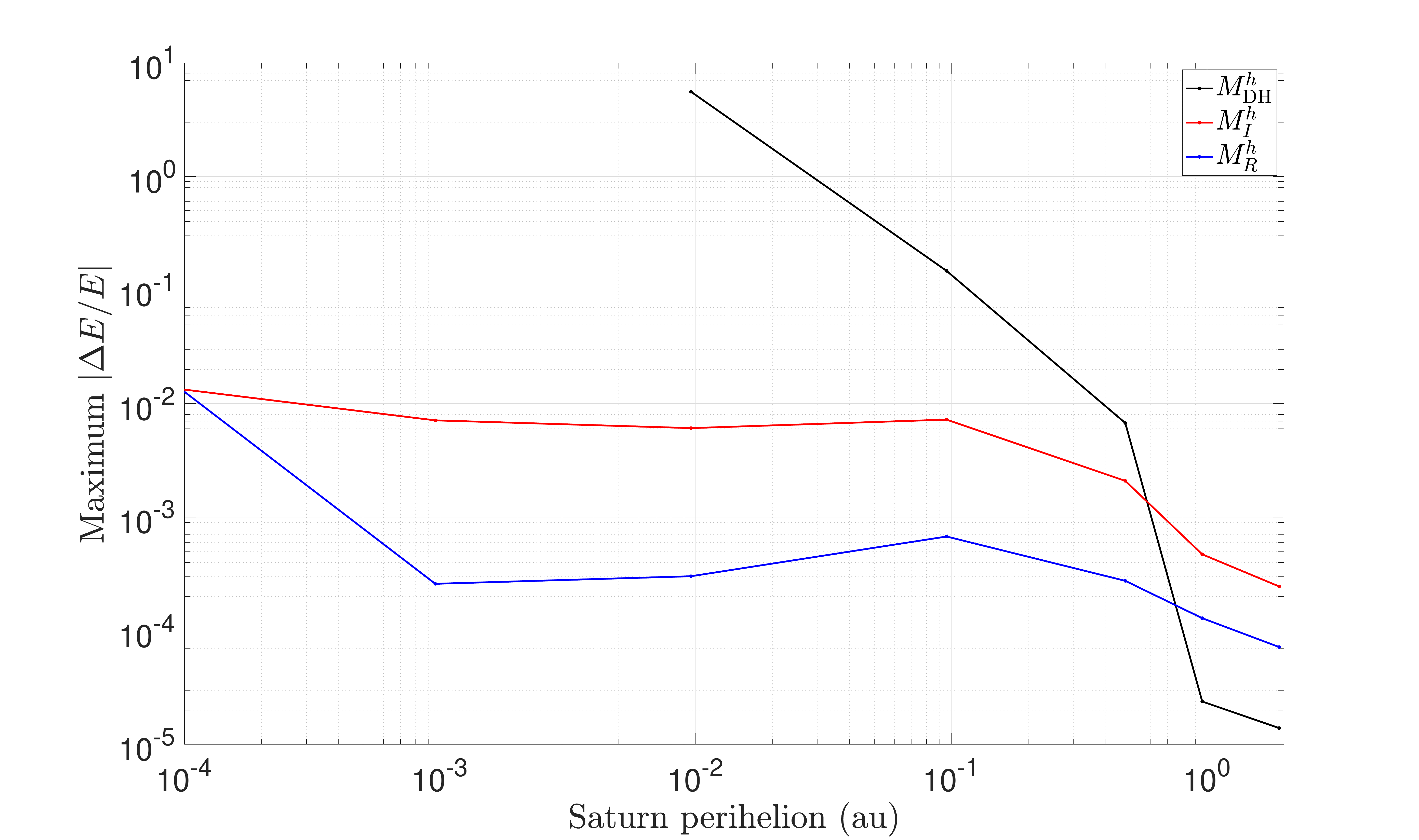}
	\caption{Maximum absolute energy error in $3000$ yr integrations of a three-body system consisting of Sun, Jupiter, and eccentric Saturn.  The error is plotted as a function of $a (1-e)$, where $a$ and $e$ are Saturn's initial semi-major axis and eccentricity, respectively.  We switch integrators discretely near Saturn's pericenter.  $M_R^h$ performs better than $M_I^h$, as usual.  The behavior of $M_R^h$ is comparable to, and even outperforms the smooth and more computationally complicated algorithm of \protect\cite{LD00}, as can be seen by comparing to their Fig. 1.  $M_{\mathrm{DH}}^h$ is simply the Wisdom--Holman map in Democratic Heliocentric coordinates.
	\label{fig:ld}
  	}
\end{figure}

In Fig. \ref{fig:ld}, we plot the maximum of the absolute energy error of the integration as a function of Saturn's initial perihelion, $a(1-e)$, where $a$ and $e$ are initial conditions.  
As usual, $M_R^h$ performs significantly better, typically an order magnitude better, than $M_I^h$.  It only required redoing between $4.6\times 10^{-3}$ and $5.2 \times 10^{-3}$ steps.  Compared to Fig. 1 in \cite{LD00}, $M_I^h$ has worse energy errors than their method labelled ``Modified DH.''  However, $M_R^h$ is comparable to, and even outperforms their algorithm at times despite our more stringent switch criteria.  So we have achieved similar error performance with simpler and fewer calculations.  We have also plotted $M_{\mathrm{DH}}^h$, in which $K = 0$ for all time.  It closely follows the curve  ``DH'' in \cite{LD00}, as expected.

Combining the ideas of Secions \ref{sec:merc} and \ref{sec:ld}, we can design a new hybrid integrator which simply and reversibly integrates close encounters between non-Solar planets and close encounters with the Sun.

\section{Conclusion}
\label{sec:conc}
We presented an algorithm for adaptive $N$-body simulations that is almost time-reversible.  The method switches discretely between two or more time steps or integrators to resolve small scale dynamics when necessary.  This algorithm presents no computational burden in our tests while improving errors by up to several orders of magnitude simply by redoing steps which violated the time-symmetric selection condition, which constitute a fraction of all time steps.  

We first presented tests that integrate Kepler and simple harmonic oscillator Hamiltonians adaptively.  We then presented realistic few-body $N$-body simulations from planetary dynamics: first we studied a chaotic exchange orbit of the restricted three-body problem, using a modified \texttt{Mercury} code.  We are able to significantly simplify it by making the switches discrete and reversible.  We can also simply incorporate velocity dependent switching functions into this code, which we have not seen done before.  Then we solved a planetary system with an eccentric planet using adaptive step sizes or switching integrator techniques, significantly simplifying the approach of \cite{LD00}.  Further applications of our algorithm are easy to envision; for example, \texttt{PETAR} \citep{Wangetal2020} is a code for simulating stellar cluster dynamics with applications such as gravitational wave predictions.  The code uses a smooth switching function as in Section \ref{sec:merc} for resolving small scale dynamics like compact binaries which can be made reversibly discrete via our techniques.  We plan to apply the ideas in this work to this code, and we plan to explore how to adapt these concepts to integrators in cosmological simulations.

As it stands, the method is only applicable to integrations using a global time step. Such methods traditionally are symplectic with a constant timestep, but inefficient. For efficiency the constant timestep must be avoided, which also renders the method non-symplectic. Then instead of symplecticity, reversibility still offers excellent error properties. This can be achieved in two ways:
\begin{itemize}
\item
using a continuously adaptive time step \citep{Handsetal2019,Boekholtetal2022}
\item
switching between integrators (the method in this paper).
\end{itemize}

Our algorithms are presented in Listings~\ref{lst:simpleAnd} and \ref{lst:revers:eff} and are easy to implement: the example comprises a few code lines.  If the algorithm switches between more than two integrators or steps, the number of conditionals and algorithm will grow in complexity, but without necessarily affecting computational burden.  Because this algorithm can lead to such significant error improvements without incurring costs, we hope it can be widely useful.

\section{Acknowledgements}
Comments by Uddipan Banik, Matt Holman, Greg Laughlin, Jun Makino, Ander Murua, and Frank C. van den Bosch improved this work.  DMH acknowledges support from the CycloAstro project.

\section{Data Availability}
The data underlying this article will be shared on reasonable request to the corresponding author.
\bibliographystyle{mnras}
\bibliography{paper}

\appendix
{
\section{A general almost reversible algorithm}
\label{app:algo}
The algorithms presented in Section~\ref{sec:algo} use combinations of $F(y_0)$ and $F(y_1)$ as conditions for using $M_1^h$. This does not allow the use of, for example, the change of acceleration over one time step. To allow that, we now consider a reversible condition $C(y_0,y_1)$ (a function returning a boolean), i.e.\ one that satisfies $C(y_0,y_1) = C(y_1,y_0)$ and
\begin{equation}
\label{eq:Crev}
    C(y_0, y_1) = C(\hat \rho y_0, \hat \rho y_1).
\end{equation}
An algorithm implementing this condition is given in Listing~\ref{lst:revers}.}%
\begin{lstlisting}[label={lst:revers},caption={\texttt{Python} code for an algorithm using condition $C(y_0,y_1)$.},
basicstyle=\ttfamily\footnotesize,
breaklines,
language=Python]
01 def mapReversibleC(y0,Mtry,M1,M2,C):
02     ytry = Mtry(y0)
03     Malt = M1 if C(y0,ytry) else M2
04     if Malt == Mtry:
05         return ytry, Mtry
06     yalt = Malt(y0)
07     if Malt == M2  or  C(y0,yalt):
08         return yalt, Malt
09     return ytry, Mtry
\end{lstlisting}
{
Here, the input $M_{\mathrm{try}}^h$ is the map, either $M_1^h$ or $M_2^h$, used in the previous step and is attempted first in line \texttt{02}. If the resulting step was consistent with the condition $C(y_0,y_1)$ tested in line \texttt{04}, it is accepted. Otherwise, the other integrator, $M_{\mathrm{alt}}^h$, is attempted in line \texttt{06} and used if consistent with the condition $C$. Finally, if neither integrator is consistent, $M_2^h$ is used (these last two conditional clauses are combined into one in line \texttt{07}).
}

\section{Hamiltonians for the simple harmonic oscillator}
\subsection{The symplectic Euler Hamiltonian}
\label{sec:DKH}
We compute the Hamiltonian obeyed by $M_{\mathrm{DK}}^h$ applied to the SHO problem of Section \ref{sec:SHO}.  From inspection of the Baker-Campbell-Hausdorff series  \citep[Section III.4.2]{hair06}, $M_{\mathrm{DK}}^h$ obeys Hamiltonian,
\begin{equation}
\label{eq:DK}
H_{\mathrm{DK}} = a(h) \left[A+B+\tfrac12 h(\v{q} \cdot \v{p})\right],
\end{equation}
We find the specific form of $a(h)$ by solving the system of homogeneous linear differential equations from Hamiltonian \eqref{eq:DK}.  Then, we compare the solution at time $h$ to $y_1 = M_{\mathrm{DK}}^h y_0$.  We have,
\begin{equation}
\begin{aligned}
q_{x,1} &= q_{x,0} + h p_{x,0} ~~~~\mathrm{, and}\\
p_{x,1} &= p_{x,0} (-h) + p_{x,0} (1 - h^2),
\end{aligned}
\end{equation}
and similar for $q_{y,1}$ and $p_{y,1}$.  We find then that $a(h) = \arcsin(X)/X$, where $X = h \sqrt{1 - h^2/4}$. Note Hamiltonian \eqref{eq:DK} is unphysical for $h > 2$.
For the reverse, $M_{\mathrm{KD}}^h = \mathrm{e}^{h \hat{A}} \mathrm{e}^{h \hat{B}}$, 
\begin{equation}
H_{\mathrm{DK}} = a(h) \left[A+B-\tfrac12 h(\v{q} \cdot \v{p})\right],
\end{equation}
which differs by a sign.

\subsection{The leapfrog Hamiltonian}
\label{sec:DKDH}
We compute the Hamiltonian obeyed by $M_{\mathrm{DKD}}^h$ applied to the SHO problem of Section \ref{sec:SHO}.  Now note $(\chi^{h})^{-1}  M_{\mathrm{DKD}}^h \chi^{h} = M_{\mathrm{KD}}^h$, where $\chi^h = \mathrm{e}^{\frac{h}{2} \hat{A}}$.  This implies the $M_{\mathrm{KD}}^h$ and $M_{\mathrm{DKD}}^h$ maps are conjugate via the map $(\v{q},\v{p}) \rightarrow (\v{q} + h/2 \v{p},\v{p})$ to the initial conditions.  From this, we derive the conserved Hamiltonian for $M_{\mathrm{DKD}}^h$:
\begin{equation}
H_{\mathrm{DKD}} = a(h) \left[\left(1- \tfrac14h^2\right) A + B\right],
\label{eq:DKD}
\end{equation}
which is easy to verify numerically.  That it is second order in $h$ can be seen immediately. $H_{\mathrm{DKD}}$ can be written in closed form for Hamiltonian \eqref{eq:HAB}, \eqref{eq:Asho}, and \eqref{eq:Bsho}, but it cannot if we are solving the Kepler Hamiltonian.

\end{document}